\documentclass[aps,pre,twocolumn,groupedaddress,floatfix]{revtex4}

\usepackage{graphicx,bbold}
\usepackage{color}
\usepackage{natbib}

\newcommand{\op}[1]{\hat{\rm{#1}}}

\begin{document}

%\preprint{}

\title[Matrix algorithm for Schr\"odinger equations]{Matrix algorithm for solving 
%Hermitian or non-Hermitian 
Schr\"odinger equations with position-dependent mass or complex 
optical potentials}

\author{Johann F\"orster, Alejandro Saenz and Ulli Wolff}
\address{Institut f\"ur Physik, Humboldt-Universit\"at zu Berlin,
        Newtonstr.~15, 12489 Berlin, Germany}

%\date{\today}

%\email[x]{Your e-mail address}
%\thanks{}
%\affiliation{}


\begin{abstract}
We represent low dimensional quantum mechanical Hamiltonians
by moderately sized finite matrices that reproduce the lowest
O(10) boundstate energies and wave functions to machine precision.
The method extends also to Hamiltonians 
that are neither Hermitian nor PT symmetric and thus allows
to investigate whether or not the spectra in such cases are still real. 
Furthermore, the 
approach is especially useful for problems in which a 
position-dependent mass is adopted, for example in effective-mass 
models in solid-state physics or in the approximate treatment of 
coupled nuclear motion in molecular physics or quantum chemistry. 
The performance of the algorithm is demonstrated by considering 
the inversion motion of different isotopes of ammonia molecules 
within a position-dependent mass model and some other examples of  
one- and two-dimensional Hamiltonians that allow for the comparison 
to analytical or numerical results in the literature.
\end{abstract}

%\pacs{}
%\keywords{}

\maketitle

\section{Introduction}
In a number of cases complex systems can be treated in a 
simplified manner, if a position-dependent mass is introduced. 
One prominent example is the concept of an effective electron mass 
in solid-state physics where position-dependent masses provide 
a way to obtain corrections to the simplest approach 
in which a constant effective mass is used (see, e.\,g., 
\cite{gen:leva10} and references therein). As a consequence, 
Schr\"odinger equations with a position-dependent mass 
have been considered in various contexts, for example to study 
electronic properties of semiconductors \cite{gen:bast88}, He clusters 
\cite{gen:barr97b}, or super-lattice band structures \cite{gen:bast81}. 

Also in molecular physics or theoretical chemistry position-dependent 
masses can occur, if high-dimensional nuclear motion is described 
using a lower dimensional effective Hamiltonian. A rather well known 
example is the theoretical description of the 
inversion motion (umbrella mode) in ammonia molecules (NH$_3$) in 
which the nitrogen atom is moving from one side of the plane formed by the 
three hydrogen atoms to the other. In fact, this involves 
a collective motion, since the hydrogen-nitrogen bonds change their 
lengths while the nitrogen atom moves. This is thus an example of 
strongly coupled vibrational modes, in this case of the symmetric 
bending and stretching modes. While the main physics of this 
motion can be captured by an effective one-dimensional double-well 
potential, an improved approximation is achieved by introducing a 
position-dependent mass while still maintaining a one-dimensional treatment. 
The inversion motion of ammonia and its 
one-dimensional model has been studied extensively in several 
papers, e.\,g., in \cite{gen:mann35,gen:rush97,gen:aqui98,gen:klop01}, 
and has been exploited for the ammonia maser \cite{gen:gord54}.
Especially Aquino $\it{et}$ $\it{al.}$\ \cite{gen:aqui98} obtained very 
accurate results in comparison to experiment by reducing
the two-dimensional problem to a one-dimensional 
problem with a position-dependent reduced mass.
Note that this ansatz is deduced from classical arguments
and thus leads to quantum mechanical operator ordering ambiguities, since
it is not clear anymore how to order the operators in the kinetic term
$p^2/2m$.
The accurate and efficient solution of a Schr\"odinger 
equation with a position-dependent mass is non-trivial and 
various efforts were made even recently to find 
analytical or numerical solutions, e.\,g., in 
\cite{gen:kill11, gen:leva10, gen:jha11}.

In this work a matrix method is introduced that determines 
efficiently and accurately the bound states of  
Schr\"odinger equations with a position-dependent mass. The algorithm 
is very flexible and allows to consider different symmetrized or 
non-symmetrized forms of the kinetic-energy operator. Furthermore, also 
non-Hermitian Hamiltonians can be treated, including complex ones. 
Non-Hermitian PT-symmetric Hamiltonians have recently stirred some interest
\cite{gen:bend98,gen:most02} in connection with the question under which 
circumstances the spectrum may still remain real. In addition to that, non-Hermitian 
Hamiltonians (also without PT-symmetry) occur for example in the context of complex
optical potentials (see, e.\,g., \cite{sfm:saen00c,sfm:goll06}).

After a brief discussion of Hamiltonians with a position-dependent mass or 
PT symmetry in Sec.~\ref{sec:pdm} the method is presented in 
Sec.~\ref{sec:method}. 
In Sec.~\ref{sec:examples} the performance 
of the method is discussed for a number of examples. 
This includes a study of the inversion mode of NH$_3$ 
(Sec.~\ref{sec:NH3ex}) and ND$_3$ (Sec.~\ref{sec:ND3}), 
the Morse potential (Sec.~\ref{sec:Morse}), a harmonic 
oscillator with position-dependent mass (Sec.~\ref{sec:PDMHO}), 
and examples of PT-symmetric and non-symmetric non-Hermitian 
Hamiltonians (Sec.~\ref{sec:pttest}). 
We also use the example in Sec.~\ref{sec:NH3ex} to discuss the 
convergence properties of the method (Sec.~\ref{sec:convergence}).
In addition to these one-dimensional examples, we discuss the application of the
algorithm to the two-dimensional Henon-Heiles system (Sec.~\ref{sec:HHS}).

\section{Position-dependent mass and PT-symmetric 
Schr\"odinger equations}
\label{sec:pdm}

We want to replace $m \rightarrow m(\op{x})$ in one-dimensional quantum systems and,
therefore, investigate various ways to order the operators in the kinetic-energy terms.
A whole class of Hermitian Hamiltonians with position-dependent mass is given by 
the von Roos Hamiltonians \cite{gen:roos83}, 
\begin{eqnarray}
  \hat{\rm{H}}=\frac{1}{4}\left(
       \op{m}^{\alpha}\op{p}\,\op{m}^{\beta}\op{p}\op{m}^{\gamma}
       + \op{m}^{\gamma}\op{p}\,\op{m}^{\beta}\op{p}\op{m}^{\alpha}
                          \right) + V(\op{x})
\end{eqnarray}
with
\begin{eqnarray}
  \alpha+\beta+\gamma=-1
\end{eqnarray}
and $\op{m}=m(\op{x})$ with the usual canonical operators $\op{p}$ and $\op{x}$.
The choice $\alpha=\gamma=0$, $\beta=-1$ leads to
\begin{eqnarray}
   \hat{\rm{H}}=\frac{1}{2}\op{p}\,\frac{1}{\op{m}}\,\op{p}+V(\op{x}),
                                                                   \label{eq:m04}
\end{eqnarray}
while with $\alpha=-1$, $\beta=\gamma=0$ one finds
\begin{eqnarray}
   \hat{\rm{H}} 
       =\frac{1}{4} \left(\frac{1}{\op{m}}\,\op{p}^2
                          + \op{p}^2\,\frac{1}{\op{m}}
                    \right)+V(\op{x}) . \label{eq:m02}
\end{eqnarray}
While there is no {\it a priory} reason to favor any of the 
Hermitian Hamiltonians, for example a specific choice of 
$\alpha, \beta$, and $\gamma$, non-Hermitian Hamiltonians 
are often rejected for general reasons, since they can 
possess complex eigenvalues. However, in standard 
quantum mechanics {\it closed} systems are described by Hermitian 
Hamiltonians and real eigenvalue spectra. (Note that non-Hermitian 
Hamiltonians occur, e.\,g., in the approximate description of 
open systems with optical potentials.) 
Regarding the reality of the spectrum, non-Hermitian PT-symmetric Hamiltonians 
form a special class as discussed in \cite{gen:bend98,gen:most02}. 
Therefore, they are of special interest for possible extensions of standard 
quantum mechanics. Note, however, that these extensions often also 
involve a complex potential instead of a position-dependent mass. 
For the NH$_3$ molecule the position-dependent reduced 
mass $\mu(x)$ in \cite{gen:aqui98} was derived within classical mechanics and  
then translated into quantum mechanics in the form
\begin{eqnarray}
   \hat{\rm{H}}=\frac{1}{2\op{m}}\,\op{p}^2+V(\op{x}) \label{eq:m01}
\end{eqnarray}
which is non-Hermitian, but convenient for computation. 
The advantage of Eq.~(\ref{eq:m01}) is that no first derivative of the 
wavefunction occurs. This simplifies the numerical 
solution with standard approaches. For example, the Numerov-Cooley 
method \cite{gen:cool61,gen:john77} can be adopted as was done in 
\cite{gen:rush97}. Another possible non-Hermitian choice is
\begin{eqnarray}
    \hat{\rm{H}}=\op{p}^2\frac{1}{2\op{m}}+V(\op{x}) . \label{eq:m03}
\end{eqnarray}
(Note if both the potential $V$ and the mass $m$ possess inversion symmetry, 
$V(x)=V(-x)$ and $m(x)=m(-x)$, as is the case for ammonia, 
then the Hamiltonians in Eqs.~(\ref{eq:m01}) and (\ref{eq:m03}) 
are P and T symmetric.)

It is important to not only have an efficient and reliable solver 
for one version of the Hamiltonian, but to be able to compare 
the results of different versions. Ideally, the results agree 
sufficiently well and thus the question of the proper form 
becomes practically inessential for the attempted approximation level.
A strong variation of the 
results with the chosen form of the Hamiltonian, on the other hand, 
is a clear 
warning signal. Since non-Hermitian and (for example for 
$V(x)\neq V(-x)$) in principle even 
PT non-symmetric Hamiltonians can be obtained for some 
versions of the Hamiltonians with a position-dependent mass, 
it is also of interest to have a solver that can handle these 
cases and that detects possible non-vanishing imaginary components 
of the eigenvalues. This is, in fact, also an important issue 
in the general context of PT-symmetric Hamiltonians (or other 
proposals for extensions of standard quantum mechanics) without 
position-dependent masses. Since in the majority of cases 
physically relevant Hamiltonians do not possess analytical 
solutions, it is important to numerically check whether the 
eigenvalue spectrum is purely real or not. The algorithm presented 
below provides a promising solution to this type of problems. 
Position-dependent masses are easily handled, even for different 
formulations of the kinetic-energy operator. Furthermore, 
a number of bound states are obtained simultaneously within 
a single calculation. Finally, the algorithm can equally well 
be applied to complex Hamiltonians and thus also to non-Hermitian 
Hamiltonians with either purely real or partly complex 
eigenvalues.

\section{The Matrix Algorithm}
\label{sec:method}
The technique to investigate simple quantum systems reviewed below has 
been taught by one of us (U.~W.)
in the computational physics courses at Humboldt university
since 2006. It has also recently been used for supersymmetric quantum mechanics
in \cite{Wozar:2011gu}. The basis is the so-called SLAC derivative
that was proposed for lattice fermions in \cite{Drell:1976mj}. While it had
to be discarded for four dimensional quantum field theory due to
incompatibilities with ultraviolet renormalization \cite{Karsten:1979wh} no such problems seem
to hamper the present quantum mechanical applications.

We define a one-dimensional position space lattice consisting of
an odd number of $N=2M+1$ points
\begin{eqnarray}
\{x_1,x_2,\ldots, x_N\}= \nonumber\\
\{-Ma,-(M-1)a, \ldots, -a,0,a,\ldots, Ma\}.
\end{eqnarray}
that are equidistantly spaced with the lattice spacing $a$
and lie symmetrically around the origin. 
Wave `functions' in the Schr\"odinger representation
of quantum mechanics
are restricted to this space 
\begin{eqnarray}
\Psi(x)\equiv \left(\begin {array} {c} \Psi(x_1) \\ 
    \Psi(x_2) \\ \vdots \\ \Psi(x_N) \end {array} \right),
\end{eqnarray}
and thus the Hilbert space is approximated by $\mathbb{C}^N$.
Intuitively, a bound state that is centered around the origin,
can be represented well in this framework, if its size is much smaller
than $L=Na$ such that the sites $\pm Ma$ are deeply in the classically forbidden region.
Moreover $a$ must be small enough to allow for a good resolution of structures
in the wave function. For example, for a simple harmonic oscillator of mass
$m$ and frequency $\omega$ these conditions amount to $a \ll \sqrt{\hbar/(m\omega)} \ll L$.

Linear operators must become finite matrices now. Functions of the position
operator like the potential $V(\op{x})$ trivially translate into diagonal matrices 
\begin{eqnarray}
 (V\Psi)(x) \equiv \left(\begin {array} {c} V(x_1)\Psi(x_1) \\
    V(x_2)\Psi(x_2) \\ \vdots \\ V(x_N)\Psi(x_N) \end {array} \right)  .
\end{eqnarray}

To also implement a canonically conjugate momentum operator $\op{p}$
that has to mimic the derivative, we need to impose boundary conditions which we take
periodic with period $L$. Note that this also implies a periodic extension
of the diagonal elements of the position operator. This means that odd powers of 
$\op{x}$ create jumps at odd-integer multiples of $L/2$, in particular at $\pm Ma$,
which will, however, be seen to cause no problems in our applications here.
The first idea that comes to mind now is to represent $\op{p}^2$ by
a difference operator over three (or more) sites. This would clearly inflict leading
discretization errors that are powers of $a^2$.
We found that a better precision is obtained by Fourier transforming, then using a
diagonal operator for $\op{p}^2$ analogous to the one in discrete position space
and then transforming back. This results in a nonlocal matrix representation 
in position space
that couples {\em all} sites given by
\begin{eqnarray}
(f(\op{p}))_{ik}=\frac{1}{N}\sum_{p} f(p) \exp[ip(x_i-x_k)].
\end{eqnarray}
where the $p$-sum runs over the values
\begin{eqnarray}
\{p_1,p_2,\ldots, p_N\}= \nonumber\\
\left\{-M\frac{2\pi}{L},-(M-1)\frac{2\pi}{L}, \ldots,
 M\frac{2\pi}{L}\right\}
\end{eqnarray}
and periodicity under shifts of $p$ by $\pm 2\pi/a$ holds.

A multiplication of a wave function with the above matrix
may be decomposed into two steps. First we compute
\begin{eqnarray}
\tilde{\Psi}(p) = a\sum_{k=1}^{N} \exp[-ip x_k] \Psi(x_k) 
\to \nonumber\\
 \int_{-L/2}^{L/2} dy \exp[-ip y] \Psi(y)
\end{eqnarray}
where the continuum limit $a\to 0$ at fixed $L$ is indicated
(for a continuously defined $\Psi$).
In the second step we form
\begin{eqnarray}
(f(\op{p})\Psi) (x)=\frac1{L} \sum_p f(p)\exp[ip x] \tilde{\Psi}(p).
\end{eqnarray}
Here $x$ can in any case assume continuous values and inside $f$
obviously $p$ is equivalent to $-i d/dx$. The $p$-sum becomes infinite
in the continuum limit.

Returning to finite $a$ and $L$
the formula (setting $j=i-k$)
\begin{eqnarray}
(\exp[i\alpha\op{p}])_{ik} &=&\frac1{N} \sum_p \exp[ip(\alpha+x_i-x_k)] \nonumber\\
&=&\frac{(-1)^j}{N}\; \frac{\sin(\pi\alpha/a)}{\sin(\pi (\alpha+ja)/L)}
\end{eqnarray}
which holds for general $\alpha$,
allows for the explicit computation of matrix elements of powers of $\op{p}$.
Comparing powers in $\alpha$ we find for example 
\begin{eqnarray}
(\op{p})_{ik}&=&\left\{
\begin{array}{ccc}
0&\mbox{for} & j=0 \mbox{ mod } N\\[1ex]
\frac{\pi}{iL}\frac{(-1)^{j}}{\sin(\pi j/N)}&\mbox{else} & \\
\end{array}
\right. ,\label{eq:00}\\[1ex]
(\op{p}^2)_{ik}&=&\left\{
\begin{array}{ccc}
 \frac{\pi^2}{3 a^2}(1-a^2/L^2) &\mbox{for} & j=0 \mbox{ mod } N\\[1ex]
\frac{2\pi^2}{L^2}\frac{(-1)^{j}\cos(\pi j/N)}{\sin^2(\pi j/N)}&\mbox{else} & \\
\end{array}
\right. .
\label{eq:01}
\end{eqnarray}
Equivalently, we may of course also take matrix powers of $(\op{p})_{ik}$.

To find the eigenvalues of the Schr\"odinger equation
we now construct the
matrix representation of $\hat{\rm{H}}$ and 
diagonalize this matrix to find
eigenenergies $E_n$ and eigenfunctions 
$\Psi_n$ of the system.
Clearly, the width $L$ and the number of points 
$N$ should be sufficiently large to obtain converged 
results and only a certain number of low-lying energies $E_n$ can
be expected to be reliable for any fixed $N$ and $L$.

To diagonalize the matrix, we implement the matrix
in MATLAB \cite{gen:matl} and employ the routine 
{\it eig} which determines eigenvalues and eigenvectors of a 
matrix (also for non-Hermitian matrices). 
A simple MATLAB code to construct the 
matrix representation of the Hamiltonian for the first
application discussed in Sec.~\ref{sec:NH3ex} 
(ammonia inversion) is explicitly given in the appendix.

As will be discussed in Sec.~\ref{sec:convergence}, the matrix
algorithm has a very fast convergence behavior such that 
around 100 points are usually sufficient. Therefore, 
also higher dimensional problems with for example about 
$N_x \times N_y = \rm{O}(10^4)$ points in two dimensions
are within reach on present-day computers.
In that case we discretize each direction as before and obtain a rectangle
filled with sites
\begin{eqnarray}
\vec{r}_I=(x_{i_1},y_{i_2}),\quad i_1=1,\ldots,N_x, \quad i_2=1,\ldots,N_y
\end{eqnarray}
with the unique compound index
\begin{eqnarray}
I=i_1+(i_2-1)N_x=1,\ldots,N_x N_y.
\end{eqnarray}
Operators are embedded in the tensor product state space.
The position operators become for example
\begin{eqnarray}
(\op{x})_{IK}=x_{i_1} \delta_{i_1 k_1} \delta_{i_2 k_2},\quad
(\op{y})_{IK}=y_{i_2} \delta_{i_1 k_1} \delta_{i_2 k_2}.
\end{eqnarray}
Similarly $\op{p}_x^2$ leads to a matrix
\begin{eqnarray}
(\op{p}_x^2)_{IK}= (\op{p}^2)_{i_1 k_1} \delta_{i_2 k_2}, \quad
(\op{p}_y^2)_{IK}=  \delta_{i_1 k_1} (\op{p}^2)_{i_2 k_2}
\end{eqnarray}
with the matrices $(\op{p}^2)_{ik}$ taken from (\ref{eq:01}) with the obvious
substitution of $a,L,N$ by the corresponding quantities referring to the
respective direction. Note that in more than one dimension many vanishing matrix elements
appear in a typical Hamiltonian. One may thus consider to store it in a sparse matrix mode
as a list of the nonvanishing elements rather than a full matrix.

The two-dimensional Henon-Heiles system is treated as an example 
in Sec.~\ref{sec:HHS} and the MATLAB \cite{gen:matl} code which
generates the matrix representation of this two-dimensional
problem is explicitly given in the appendix.

The algorithm is not time critical for the one-dimensional
problems discussed in this paper, which means we 
find well converged results for these systems
in less than a second on a modern standard computer.
For the two-dimensional Henon-Heiles system, 
around 5 minutes are needed to find the eigenenergies
and around 60 minutes are required to find
the eigenfunctions in addition when using 101x101 points.
Therefore, an implementation in programming 
languages like FORTRAN or C which would speed up the 
algorithm is a possible future option, but was not needed 
for the examples considered in this work. 
Furthermore, with only a single matrix diagonalization many bound 
states with eigenenergies $E_n$ 
are found simultaneously.
Other algorithms \cite{gen:riva91, gen:kill11} which were previously 
used to solve the one-dimensional Schr\"odinger equation 
with position-dependent masses consist of 
``guessing'' an eigenenergy $E$ and a subsequent test   
whether the resulting wave function has the correct 
behavior to fulfill the Schr\"odinger equation or not.

\section{Example applications of the matrix algorithm}
\label{sec:examples}

\subsection{The inversion motion of NH$_3$}
\label{sec:NH3ex}

\begin{figure}
\begin{center}
    \includegraphics[width=0.5\textwidth]{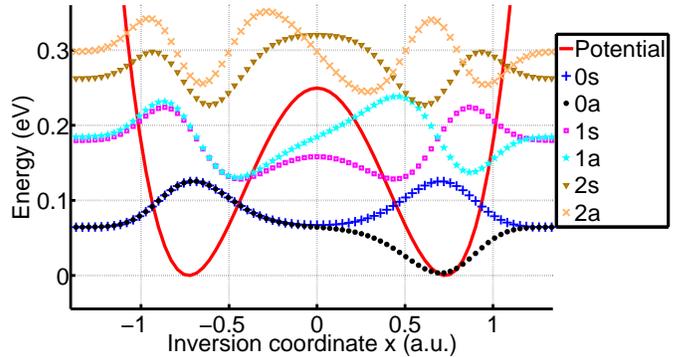}
    \caption{\label{fig:01}
    (Color online) Potential curve (solid red line, data from \cite{gen:aqui98}) 
    and wave functions (symbols, see legend in the graph) of the 6 
    energetically lowest states 0s to 2a. 
    The wave functions have been shifted by constants such that their values at 
    $\left|x\right|\rightarrow \infty$ displays the
    eigenenergies. In the calculation, 
    a width of $L=4$\,a.u.\ and $N=111$ points were used.
    }
\end{center}
\end{figure}

To calculate the energy levels describing the inversion motion 
in ammonia with our algorithm, we exactly follow the procedure in 
\cite{gen:aqui98} and first represent
the data points (from table 1 in \cite{gen:aqui98}) 
for the double-well potential $V(x)$ describing 
the inversion mode by an even polynomial of degree 20.
Our coefficients are listed in the appendix.
It might be noted that the potential obtained with DFT
is in good agreement with the experimental value for the 
equilibrium geometry in \cite{gen:swal62} and the obtained 
barrier height of 2013.5 cm$^{-1}$ agrees well with the 
one from empirical procedures (2018 $\pm 10$ cm$^{-1}$ \cite{gen:klop01}).
In addition, the theoretical study in \cite{gen:klop01} 
yields an effective barrier height of 2021 $\pm$ 20 cm$^{-1}$. 
The potential is therefore very useful for a one-dimensional 
study, though the method is nevertheless a little bit questionable, 
because the theoretical effective barrier in \cite{gen:klop01} is 
only obtained, if one takes into account the zero-point 
vibrational energies of the other vibrational degrees of
freedom, which adds a value of 244 $\pm$ 14 cm$^{-1}$ to 
the barrier height according to \cite{gen:klop01}.\\
The goal is now to solve the Schr\"odinger equation \footnote{
Unless otherwise noted, we work in atomic units with 
$\hbar=m_e=e^2=4\pi\epsilon_0=1$ which also puts $a_{\rm Bohr}$
and $E_{\rm Hartree}$ to unity. To conform with the literature
we convert some energies to 1/cm and lengths to \AA.
Conversion constants can be found in the appendix.  
},

\begin{eqnarray}
\Psi''(x)=2\mu(x)\left[V(x)-E\right]\Psi(x) ,
\end{eqnarray}
or equivalently, 
\begin{eqnarray}
\left(\frac{1}{2\mu(x)}\op{p}^2+V(x)\right)\Psi(x)=E\Psi(x)\label{eq:invsgl}
\end{eqnarray}
with the position-dependent reduced mass
\begin{eqnarray}
\mu(x)=\frac{3mM}{3m+M}+\frac{3mx^2}{r_0^2-x^2} , \label{eq:massfunc}
\end{eqnarray}
where $m$ is the mass of a hydrogen atom, $M$ the one  
of the nitrogen atom, and $r_0=1.00410198$\, a.\,u.\ is 
given in \cite{gen:aqui98} as the N-H distance 
which minimizes the energy of the molecule in planar 
geometry. The result of the matrix diagonalization
is illustrated in Fig.~\ref{fig:01} and the 
obtained energies are listed in Table \ref{tab:01}. The states are 
named using the symmetry label {\it symmetric} (s) or 
{\it antisymmetric} (a) together with an index $n$ where $n=0$ 
stands for the energetically lowest state with symmetric or 
antisymmetric character. Thus 0s designates 
the symmetric ground state and we give the other energies 
relative to this level. A comparison of the energy levels found 
with the matrix algorithm to the energy levels in table \ref{tab:01} 
of \cite{gen:aqui98} shows perfect agreement.

%\begingroup
%\squeezetable
\begin{table}
\begin{center}
\begin{tabular}{c|c|c|c|c}
State& $\mu$=const \cite{gen:aqui98}&   Ref. \cite{gen:aqui98}&     This work&    Experiment\\
\hline
0s&0.00&0.000&0.000&0.000\\
0a&1.05&0.837&0.837&0.793 \cite{gen:spir83}\\
1s&977.23&931.72&931.72&932.43 \cite{gen:gord70}\\
1a&1030.12&968.67&968.67&968.12 \cite{gen:gord70}\\
2s&1651.69&1596.76&1596.76&1598.47 \cite{gen:urba81}\\
2a&2011.44&1885.33&1885.33&1882.18 \cite{gen:urba81}\\
3s&2558.75&2389.14&2389.15&2384.17 \cite{gen:urba81}\\
3a&3142.66&2902.99&2902.99&2895.61 \cite{gen:urba81}\\
\end{tabular}
\caption{Inversion energy levels of NH$_3$ (shifted by $E_{0\rm{s}}$) in cm$^{-1}$. 
The energies of our work were found with 
$L=4.0$\,a.\,u., $N=111$ points and adopting the same 
1D potential as in \cite{{gen:aqui98}}.\label{tab:01}}
\end{center}
\end{table}
%\endgroup

As has been mentioned in Sec.~\ref{sec:pdm},
the obtained eigenenergies should ideally not change
drastically, if one puts the position-dependent
mass into the Schr\"odinger equation in another
form.
Table \ref{tab:02} shows the eigenenergies obtained with 
Eqs.~(\ref{eq:m04}) to (\ref{eq:m03}). 
We find a maximal relative deviation of the eigenenergies 
listed in Table \ref{tab:01} of about 0.5\% (0.005 cm$^{-1}$) 
for the 0a state, indicating that the choice where to 
put the reduced mass has only a very minor influence 
on the resulting energies within the given accuracy.

%\begingroup
%\squeezetable
\begin{table}
\begin{center}
\begin{tabular}{c|c|c|c|c}
State& Eq. (\ref{eq:m04}) &Eq. (\ref{eq:m02})&   Eq. (\ref{eq:m01}) & Eq. (\ref{eq:m03})\\
\hline
0s&	    0.000&	  0.000&	  0.000&    0.000\\
0a&	    0.837&	  0.833&	  0.837&    0.837\\
1s&	 	 931.71&	 932.01&	 931.72&   931.72\\
1a&	 	 968.64&	 968.81&	 968.67&   968.67\\
2s&		1596.77&	1597.36&	1596.76&  1596.76\\
2a&		1885.25&	1885.45&	1885.33&  1885.33\\
3s&		2389.03&	2389.21&	2389.15&  2389.15\\
3a&		2902.82&	2902.84&	2902.99&  2902.99\\
\end{tabular}
\caption{Inversion energy levels of NH$_3$ (shifted by $E_{0\rm{s}}$) in cm$^{-1}$
for different implementations of the position-dependent
reduced mass. In the calculations $L=4.0$\,a.\,u.\ and 
$N=111$ points were used.\label{tab:02}}
\end{center}
\end{table}
%\endgroup

\subsection{Inversion energy levels of ND$_3$}
\label{sec:ND3}

In addition to NH$_3$, we test our algorithm by finding the 
energy levels of ND$_3$ which were not calculated in \cite{gen:aqui98}. 
This is also of physical interest, since it is a further check 
of the adopted position-dependent mass model.  
Therefore, we simply replace the mass of the hydrogen atom by 
the mass of deuterium $m_{\rm{D}}$=2.013553212712 amu \cite{gen:nistC}
and solve the Schr\"odinger equation (\ref{eq:m01}).
The obtained energies are listed in Table \ref{tab:05}.
It turns out that the inversion splitting can be reproduced and
the energy levels for higher excited vibrational states have a 
relative error 
$\frac{\left|E_{\rm{calc}}-E_{\rm{exp}}\right|}{E_{\rm{exp}}}\leq 0.7\%$ 
in comparison to the experiment whereas the use of the constant reduced mass
\begin{eqnarray}
\mu=\frac{3mM}{3m+M}\left(1+\frac{3m\sin^2{\beta_e}}{M}\right) \label{eq:massconst}
\end{eqnarray}
from \cite{gen:town75} with $\beta_e = 22^{\circ}13'$ (from \cite{gen:aqui98}) 
leads to higher energy values and, therefore, larger relative errors (up to $6.6\%$).
Even though the agreement with the experiment is very good, it still has 
to be noted that according to \cite{gen:klop01} the barrier height depends on the
zero-point vibrational energy of the other vibrational degrees of freedom,
and this zero-point vibrational energy depends on the isotope.

%\begingroup
%\squeezetable
\begin{table}
\begin{center}
\begin{tabular}{c|c|c|c}
State& This work & This work&     Experiment \cite{gen:spir83}\\
&$\mu$=const Eq. (\ref{eq:massconst})&	$\mu(x)$ Eq. (\ref{eq:massfunc}) &\\
\hline
0s&	0&	0&	0\\
0a&	0.05&	0.05&	0.05\\
1s&	793.8&	746.2&	745.6\\
1a&	798.3&	749.3&	749.15\\
2s&	1419.7&	1368.4&	1359.0\\
2a&	1513.7&	1432.0&	1429.0\\
3s&	1912.6&	1836.4&	1830.0\\
3a&	2238.4&	2106.4&	2106.6\\
\end{tabular}
\caption{Inversion energy levels of ND$_3$ (shifted by $E_{0\rm{s}}$) in cm$^{-1}$. 
The calculations were done with $L=4.0$ a.u. and $N=111$ points.\label{tab:05}}
\end{center}
\end{table}
%\endgroup

\subsection{Convergence properties}
\label{sec:convergence}
\begin{figure}
\begin{center}
    \includegraphics[width=0.5\textwidth]{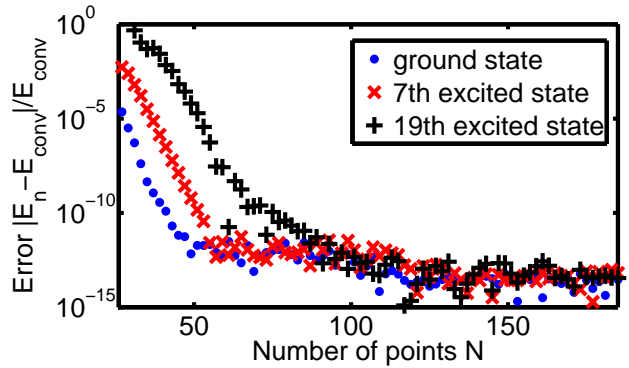}
    \caption{\label{fig:02}
    (Color online) Convergence behavior with respect to a variation of
    the lattice spacing that is proportional to  
    $1/N$ as we hold fixed the width $L=4.5$\,a.\,u. We show the ground 
    state and the 7th as well as the 19th excited state. 
    Relative errors of the energies 
    $\frac{\left|E_n-E_{\rm{conv}}\right|}{E_{\rm{conv}}}$
    are plotted,
    where $E_{\rm{conv}}$ is the `converged' energy (average over 
    the 10 largest $N$-values).
    }
\end{center}
\end{figure}

\begin{figure}
\begin{center}
    \includegraphics[width=0.5\textwidth]{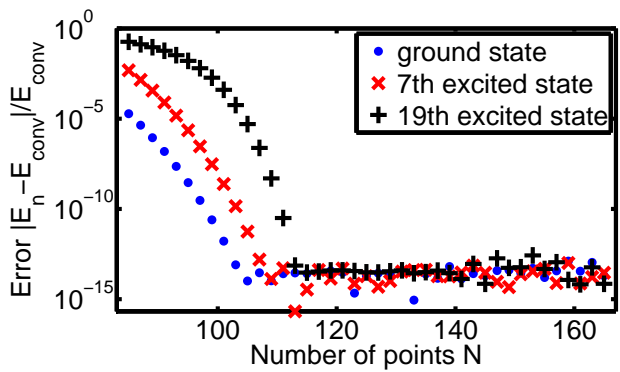}
    \caption{\label{fig:03}
    (Color online) as Fig.~\ref{fig:02}, but now $a=4.5/151$ a.~u.
    is fixed and thus
    $L$ varies proportional to the number of points $N$.
    }
\end{center}
\end{figure}

We discuss the convergence behavior of the matrix algorithm 
on the example of NH$_3$ (Sec.~\ref{sec:NH3ex}).
We use Eq. (\ref{eq:m02}) as Hamiltonian and the number 
of points $N$ is varied first for fixed $L=4.5$ a.~u.
with the result shown in Fig.~\ref{fig:02}.
In this way the dependence on the resolution
given by the lattice spacing was tested.
In Fig.~\ref{fig:03} we have frozen the spacing
to $a=4.5/151$ a.~u. and thus test stability when changing $L$.
The nearly straight lines in the semilogarithmic plot
indicate exponential convergence behavior with 
respect to both regulators.

The convergence that we have demonstrated in our first test is actually
physically plausible. As indicated before the only exponentially small
sensitivity with respect to (large) $L$ is due to the smallness of the
wave functions for bound states deep in the classically forbidden region.
The same wave functions in momentum space will also have an exponential
fall-off at large momenta so that a dual argument holds in momentum space.
We may argue here by analogy with the sampling theorem \cite{Press:1058314}.
A continuous time signal that contains no frequencies beyond the Nyquist 
frequency $\omega=\pi/\tau$ can be {\em exactly} reconstructed from sampling it
at discrete times separated by $\tau$ (CD player). In analogy, if the support
of our bound state wave functions would be {\em exactly} contained in the interval
$[-\pi/a, \pi/a]$ then the {\em exact} wave function for continuous $x$ could be
reconstructed form the Fourier components (discrete and finite in number for finite $L$). 
Then
there obviously is an exact correspondence between derivatives $d/dx$ and
factors $ip$. In reality the boundstates do not have compact momentum
support, but the deviation is only caused by the exponentially small tales
for small enough $a$.

\subsection{Morse potential}
\label{sec:Morse}
The discrete position (momentum) space introduced in the
matrix algorithm is symmetric with respect to $x=0$ ($p=0$) as 
is also the double-well problem discussed in the
previous section.
Nevertheless, general Hamiltonians can be
non-symmetric and may also have a 
partially continuous  spectrum (in the infinite volume).
Therefore we want to investigate, if bound 
states can still be found accurately and efficiently using the 
matrix algorithm also in this more general case.
A popular example for a non-symmetric potential with
bound and continuum states is the Morse potential 
\begin{eqnarray}
V(R)=D_e\left(1-e^{-\alpha(R-R_e)}\right)^2 . \label{eq:morse}
\end{eqnarray}
Often the vibration of a diatomic molecule can be 
well-described by this potential.
The energy spectrum for the bound states of 
this potential is known analytically \cite{gen:mors29}, 
\begin{eqnarray}
E_n=2\pi\hbar \nu_0
\left(n+\frac{1}{2}\right)-\frac{\pi^2\hbar^2\nu_0^2}{D_e}\left(n+\frac{1}{2}\right)^2,
\label{eq:mexact}
\end{eqnarray}
with $\nu_0=\frac{\alpha}{2\pi}\sqrt{\frac{2D_e}{\mu}}$
where $\mu$ is (again) the reduced mass. 
We investigate a Morse potential with 
parameters $D_e=\mu=1$ and $\alpha=0.24$, $R_e=-35$ 
where 6 vibrational bound states exist. 
Table \ref{tab:04} shows the energies obtained
by the matrix algorithm with $N=111$ points and $L=90$ together
with the exact results from Eq.~(\ref{eq:mexact}).
We still find excellent agreement except for some visible
discrepancy for the highest lying 6th vibrational state.
This can be understood, since states close to the continuum are 
more extended in space. If $N=201$ points, $L=140$, and $R_e=-60$ 
are used, all eigenenergies obtained by the matrix algorithm are identical 
to the analytical results within the accuracy given in Table \ref{tab:04}. 

In addition to the bound states, we obtain discretized continuum states
on which the periodic boundary conditions of the method are imprinted. 
To check whether this discretized continuum can approximately represent 
the true continuum, we examined as an example the relation
\begin{eqnarray}
\left\langle 0|\op{x}^2|0\right\rangle\approx
\sum_{n=0}^{n_{\rm{max}}}\left\langle 0|\op{x}|n\right\rangle\left\langle n|\op{x}|0\right\rangle
\end{eqnarray}
using the discretized states $\left|n\right\rangle$.
Using $N=301$ points, $L=140$, and $R_e=-60$ 
the relative error
\begin{eqnarray}
\epsilon(n_{\rm{max}})=\frac{\left|\left\langle 0|\op{x}^2|0\right\rangle-\sum\limits_{n=0}^{n_{\rm{max}}}\left\langle 0|\op{x}|n\right\rangle\left\langle n|\op{x}|0\right\rangle\right|}{\left\langle 0|\op{x}^2|0\right\rangle}\ 
\end{eqnarray}
was calculated. 
With a restriction of the summation to bound states only we find 
$\epsilon(n_{\rm max} = 5)\approx 10^{-6}$. When we further increase $n_{\rm{max}}$ 
the error decreases monotonically until machine precision is reached 
($\epsilon(n_{\rm{max}})<10^{-14}$ for $n_{\rm{max}}> 116$). 
Thus the completeness relation of the eigenstates
is numerically fulfilled, if also the discretized continuum states 
are considered. We conclude that  
any upcoming non-symmetric problem that contains a continuum spectrum
should also be well treatable by the matrix algorithm, if discretized 
continuum states (with periodic boundary conditions) are sufficient 
as is the case in the here considered example of bound to continuum 
transitions.

%\begingroup
%\squeezetable
\begin{table}
\begin{center}
\begin{tabular}{c|c|c}
$n$& Matrix algorithm&Exact\\
\hline
 0  &  0.1625056275 &  0.1625056275\\
 1  &  0.4443168825 &  0.4443168825\\
 2  &  0.6685281374 &  0.6685281374\\
 3  &  0.8351393923 &  0.8351393924\\
 4  &  0.9441506473 &  0.9441506474\\
 5  &  0.9955620565 &  0.9955619023\\
\end{tabular}
\caption{Eigenenergies of the Morse potential (\ref{eq:morse}) obtained 
with $N=111$ points and $L=90$ for the model parameters $D_e=\hbar=\mu=1$ 
and $\alpha=0.24$, $R_e=-35$. 
The exact values were obtained using Eq.~(\ref{eq:mexact}).
\label{tab:04}}
\end{center}
\end{table}
%\endgroup

\subsection{Harmonic oscillator with a position-dependent mass}
\label{sec:PDMHO}

To compare the present matrix algorithm to the recently published traditional
second-order Hartree shooting method \cite{gen:kill11}, we implemented 
the two model Hamiltonians
\begin{eqnarray}
\hat{\rm{H}}_1=\frac{1}{2}\ \op{p}\ \frac{1}{1+\op{x}^2}\ \op{p}+\frac{1}{2}\ \op{x}^2 \label{eq:h1}
\end{eqnarray}
and
\begin{eqnarray}
\hat{\rm{H}}_2=\frac{1}{2}\ \op{p} \left(\frac{1+\op{x}^2}{2+\op{x}^2}\right)^2 \op{p}+\frac{1}{2}\ \op{x}^2 .
\label{eq:h2}
\end{eqnarray}
When we compare our eigenenergies obtained by the matrix
algorithm with $N=201$ points and $L=20$ (which is well within the converged
regime), we reproduce table 1 (for $\hat{\rm{H}}_1$)
and table 3 (for $\hat{\rm{H}}_2$) 
in \cite{gen:kill11} to all digits.
Considering the convergence properties, qualitatively the same
exponential behavior is found as shown in Figs.~\ref{fig:02} and \ref{fig:03}.
One should note that the second-order shooting method in \cite{gen:kill11} only 
converges quadratically and, therefore, much larger numbers of evenly 
spaced points ($N=640,960,1440$ and $2160$) were needed to obtain the results.

\subsection{Non-Hermitian PT-symmetric and non-symmetric cases}
\label{sec:pttest}

\begin{figure}
\begin{center}
    \includegraphics[width=0.5\textwidth]{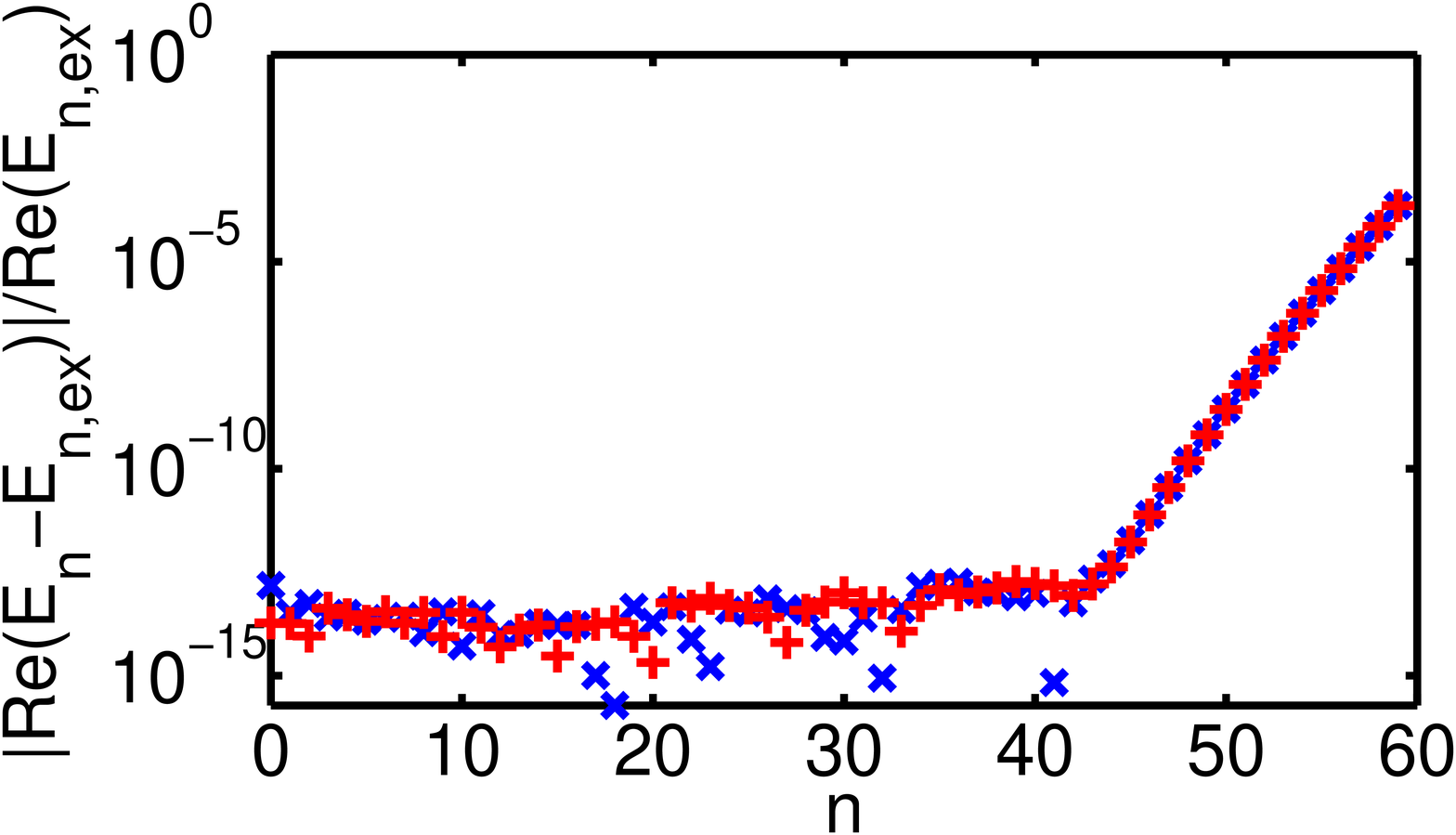}\\
    \includegraphics[width=0.5\textwidth]{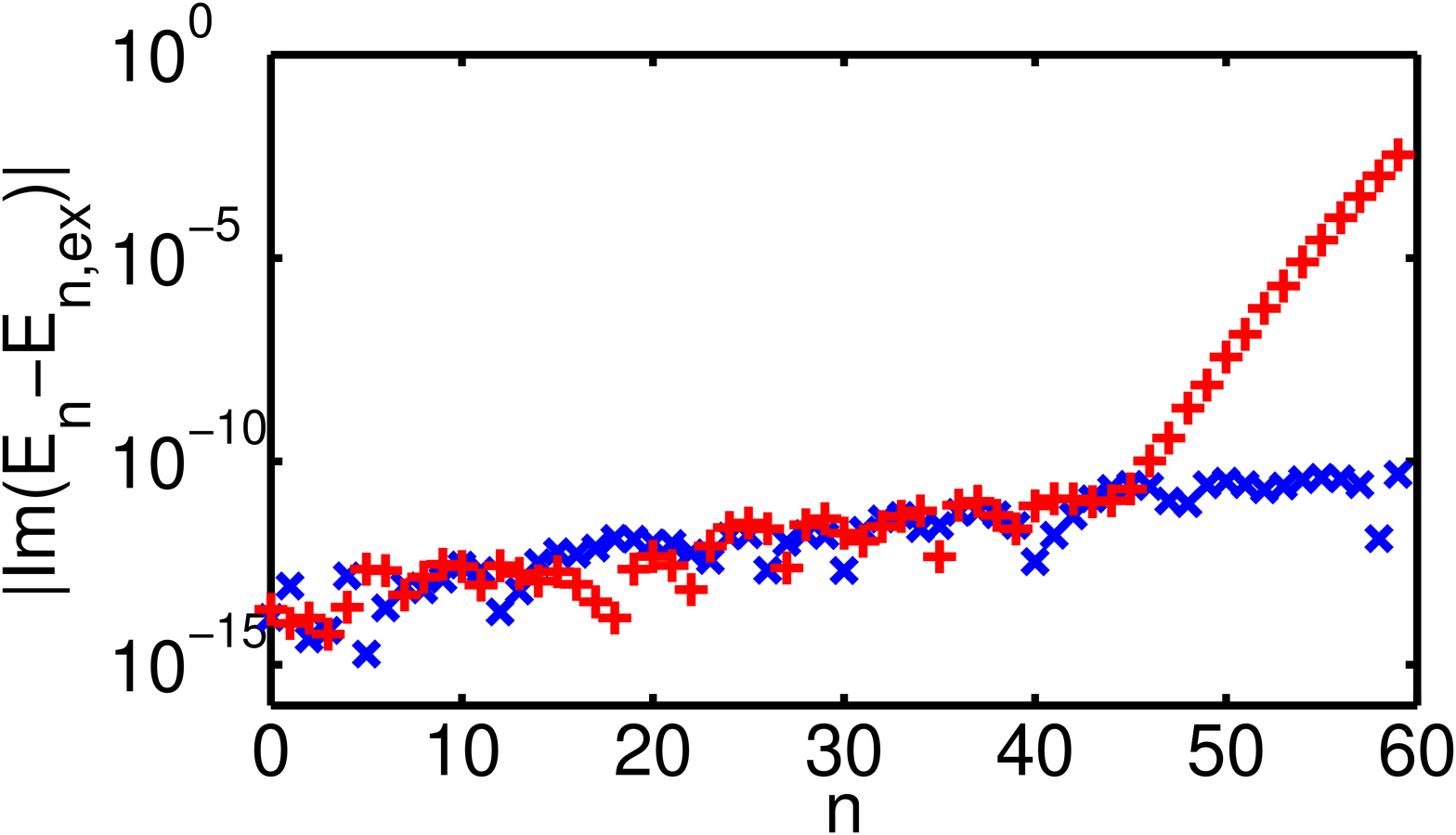}
    \caption{\label{fig:05}
    (Color online) 
    \textbf{Top}: Relative Error of the real part of the eigenvalues of
    Eqs. (\ref{eq:hcompl}) [\textcolor{blue}{x}] and (\ref{eq:hcompl2})
    [\textcolor{red}{+}].
    The values were obtained with the matrix algorithm
    using $N=101$ points and $L=25$. The exact values $E_{n,\rm{ex}}$ are given 
    by Eqs.~(\ref{eq:pteigen}) and (\ref{eq:notpteigen}).\\
    \textbf{Bottom}: Absolute Error of the imaginary part of the eigenvalues 
                    (same $L$ and $N$).
        }

\end{center}
\end{figure}

As mentioned in Sec.~\ref{sec:pdm}, the PT-symmetric Hamiltonians
examined in Sec.~\ref{sec:pdm} and \ref{sec:NH3ex} are rather special.
Therefore, we will briefly discuss the more usual form of a PT-symmetric
Hamiltonian with a constant mass but a complex potential.
We want to demonstrate that the matrix algorithm is also 
applicable for such types of problems, since complex matrices can also  
be treated. Therefore, we implemented the model Hamiltonian
\begin{eqnarray}
   \hat{\rm{H}}=\op{p}^2+\op{x}^2+i\ \op{x} . \label{eq:hcompl}
\end{eqnarray}
According to \cite{gen:bend98}, this PT-symmetric Hamiltonian has the real 
eigenenergies
\begin{eqnarray}
  E_n=2\ n+\frac{5}{4} .\label{eq:pteigen}
\end{eqnarray}
In contrast, the non-Hermitian and not PT-symmetric Hamiltonian
\begin{eqnarray}
   \hat{\rm{H}}=\op{p}^2+\op{x}^2+i\ \op{x}-\op{x} \label{eq:hcompl2}
\end{eqnarray}
has according to \cite{gen:bend98} the complex eigenvalues 
\begin{eqnarray}
    E_n=2\ n+1+\frac{1}{2}\ i \label{eq:notpteigen}
\end{eqnarray}
which can no longer be interpreted as eigenenergies. The relative error 
of the real part as well as the absolute error of the imaginary part
of the eigenvalues obtained with the matrix algorithm with $N=101$ points and $L=25$ is 
shown in Fig.~\ref{fig:05}. The eigenvalues of the Hamiltonian in Eq. (\ref{eq:hcompl}) 
now contain a small imaginary part (also see Fig.~\ref{fig:05}) due to roundoff 
errors which was not the case for the PT-symmetric
Hamiltonians in Eqs.~(\ref{eq:m01}) and (\ref{eq:m03}). The reason is that 
the matrix representations of Eqs.~(\ref{eq:m01}) and (\ref{eq:m03})
are real, while the matrix representation of Eq.~(\ref{eq:hcompl}) is complex.
Nevertheless, the very good agreement of our results with Eqs.~(\ref{eq:pteigen}) 
and ~(\ref{eq:notpteigen}) (relative error below $10^{-12}$ for $n<45$) shows 
that our matrix algorithm can also be very helpful to find eigenenergies of 
PT-symmetric Hamiltonians. 
In critical cases one could in principle consider increasing the numerical 
precision (either in steps from single to quadruple precision or, even better, 
digit-by-digit) to probe further whether the eigenvalue spectrum is purely real 
or not.

\subsection{The two-dimensional Henon-Heiles system}
\label{sec:HHS}
\begin{figure}
\begin{center}
    \includegraphics[width=0.5\textwidth]{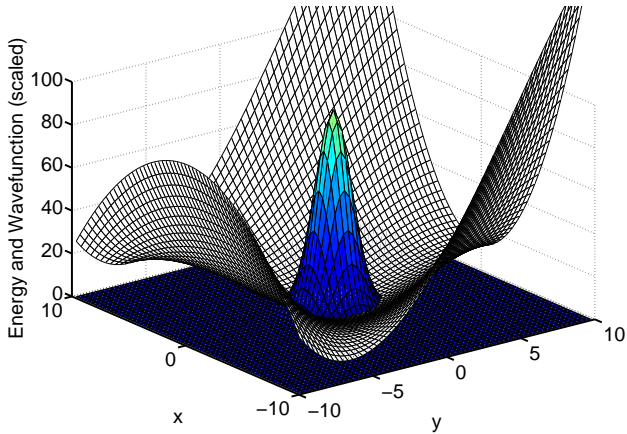}
    \caption{\label{fig:04}
    (Color online) Potential energy surface (filled white) and obtained (discrete) 
    ground state wavefunction (filled blue, scaled arbitrary)
    of the Henon-Heiles system in Eq.~(\ref{eq:2DHHS}) with $\lambda=1/\sqrt{80}$. 
    A $L_x\times L_y=20 \times 20$ grid with $N_x \times N_y=61 \times 61$ points was used.
        }
\end{center}
\end{figure}

To demonstrate the performance of the matrix algorithm in two dimensions we 
investigate the Henon-Heiles system \cite{gen:heno64}
\begin{eqnarray}
\hat{\rm{H}}=\frac{1}{2}\left(\op{p}_x^2+\op{p}_y^2\right)+\frac{1}{2}\left(\op{x}^2+
\op{y}^2\right)+\lambda\left(\op{x}^2\ \op{y}-\frac{\op{y}^3}{3}\right) .\label{eq:2DHHS}
\end{eqnarray}
This model potential is frequently used as a benchmark for numerical methods
\cite{gen:juli92,gen:zhan97,gen:poir99}, although the potential in
Eq.~(\ref{eq:2DHHS}) is not bounded from below and thus it does not support 
true bound states but only metastable ones that decay by tunneling through 
the barriers as is discussed in, e.\,g., Refs.~\cite{gen:wait81,gen:poir99}.
We use $\lambda=1/\sqrt{80}$ to compare our lowest 36 eigenenergies with the 
ones obtained in \cite{gen:poir99}.
The MATLAB \cite{gen:matl} program which generates the matrix
representation of Eq.~(\ref{eq:2DHHS}) is given in the appendix. When using a
$L_x\times L_y=20\times 20$ grid with $N_x\times N_y=61 \times 61$ points, we
reproduce all 36 eigenenergies given in \cite{gen:poir99} within the given
accuracy.
The obtained ground state wavefunction is shown together with the potential 
in Fig.~\ref{fig:04}.
When increasing the number of grid points to $N_x\times N_y=81 \times 81$ and
$N_x\times N_y=101 \times 101$ ($L_x\times L_y=20\times 20$ held constant) or
changing the grid size to $L_x \times L_y = 18 \times 18$ and $L_x \times L_y
= 22 \times 22$ ($a_x=a_y=20/61$ approximately held constant), the obtained
energies are stable within an accuracy of at least 12 significant 
digits \footnote{For the larger grid with $L_x \times L_y = 22 \times 22$ one 
additional state appears within the lowest 37 eigenenergies that is ignored 
in the comparison since its energy depends on the box size. This state is 
localized outside the potential well and its appearance is a consequence of 
the mentioned fact that the Henon-Heiles potential is not bounded from below.}.
Thus the matrix algorithm is also very suitable for finding bound states
of two-dimensional Hamiltonians.

\section{Summary}
\label{sec:summary}

The matrix algorithm presented in this paper is easy to 
implement, flexible, and shows exponential convergence  
(with respect to the number of grid points $N$ and width 
$L$). While other algorithms \cite{gen:kill11,gen:riva91} usually consist 
of guessing an initial energy, this algorithm represents a much 
more direct method, since one finds a set of  
eigenenergies by just a single matrix diagonalization.
To demonstrate the performance of the algorithm it was applied 
to a one-dimensional model describing the inversion motion
of NH$_3$ and ND$_3$. Furthermore, different forms for the 
position-dependent mass Hamiltonian were discussed. A great 
advantage is the flexibility of the present algorithm to handle 
their different possible forms. Especially, it is possible to avoid 
non-Hermitian model Hamiltonians (as used for, e.\,g., 
describing inversion of NH$_3$) that are often only adopted, 
because the symmetrized Hermitian form of the Hamiltonian leads 
to a Schr\"odinger equation that cannot be solved with many of 
the standard numerical algorithms. 

In the case of ammonia it turns out that the eigenenergies 
do not strongly depend on the choice where one puts 
the mass into the Hamiltonian. This justifies the underlying 
classical derivation of the model Hamiltonian with a position-dependent 
mass. Also the eigenenergies of a Morse potential were determined.  
The results indicate that the algorithm is applicable to find 
bound states of general Hamiltonians that may contain asymmetric 
potentials and include a continuous spectrum. The algorithm was 
also successfully applied to non-Hermitian Hamiltonians both 
with or without PT symmetry. This allows to numerically check 
whether the eigenvalue spectrum of such an Hamiltonian is purely 
real or not and demonstrates that the algorithm can handle 
Schr\"odinger equations with complex eigenvalue spectra. Finally, 
it was demonstrated with the aid of a two-dimensional problem 
that the algorithm is straightforwardly applied to higher-dimensional 
problems, only limited by the increasing numerical efforts due to the 
exploding number of grid points needed.

For very large problems or
to go beyond two dimensions it may be useful to write a routine
that applies $\op{H}$ to a vector $\Psi$ and renounce
at storing any matrix at all. Such a routine would apply FFT twice for each 
direction in every call with the cost scaling like $N^D \ln N$.
If only some low-lying energies are requested, the use of a 
Lanczos- or Arnoldi-type algorithm would be a natural choice for 
enhancing the efficiency. Within Matlab the eigenvalue finder {\tt eigs}
would allow for such an approach and {\tt fft} is available as well.

\begin{acknowledgments}
We thank Dr. Oliver B{\"a}r for helpful discussions.
\end{acknowledgments}

\begin{appendix} 
\section{Matlab code fragments}
%\section{Matlab codes}
The following short MATLAB \cite{gen:matl} 
function generates the lattice and the 
real antisymmetric matrix
$i\op{p}$ for given integer $M$ 
and width $L$ that will enter into the construction of Hamiltonians:
\begin{small}
\begin{verbatim}
function [x,ip_op] = gen_mommatrix(L,M)
% generate the lattice x      (vector)
% and i times p-operator (real,antisymmetric)
%
N=2*M+1;
a=L/N;               % spacing
x=a*(-M:M);          % x-values
ip_op=zeros(N,N);    % matrix p
c1=pi/L;
c2=pi*(N+1)/N;
for i=1:N-1
  for k=i+1:N
      ip_op(i,k)=c1/sin(c2*(i-k));
      ip_op(k,i)=-ip_op(i,k);
  end
end
\end{verbatim}
\end{small}
For the ammonia inversion problem
the following constants are required:
\begin{small}
\begin{verbatim}
%constants from http://physics.nist.gov/cuu/
Hartreecm=219474.63137;%1 a.u. (Hartree energy) 
                       %in 1/cm
aBohr=0.52917721092;   %1 a.u. (Bohr radius) 
                       %in Angstrom
%constants from ref. Aquino et al.
m=1.007825035;      %hydrogen mass in amu
%m=2.013553212712;  %deuterium mass in amu (from nist)
M=14.003074;        %nitrogen mass in amu
amu=1822.888;       %1 amu in a.u.
r0=1.00410198/aBohr;%N-H-distance in a.u.
% Potential fit parameters: 
%V=sum_{i=0}^{i=10} K(i+1) x_^{2i}
% [x in Angstrom, V in a.u.]
K=[ 0                ...
-1.2760373471398e-01 ...
 4.7973549262032e-01 ...
-4.4967805753691e-01 ...
 3.4048981035460e+00 ...
-2.5268066877745e+01 ...
 1.1565093681631e+02 ...
-3.2323821164423e+02 ...
 5.4331165379878e+02 ...
-5.0630533518111e+02 ...
 2.0128292638493e+02 ];
\end{verbatim}
\end{small}
The Hamiltonian is constructed by the following sequence of
steps (for $L=4, N=111$):
\begin{small}
\begin{verbatim}
[x,ip_op] = gen_mommatrix(4,55);
% x-dependent mass
mux=3*m*M/(3*m+M)+3*m*x.^2./(r0^2-x.^2);
mux=mux*amu; % conversion amu -> a.u.
% kinetic part
psq=-ip_op^2;
H=-0.5*ip_op*diag(1./mux)*ip_op;            %Eq.(3)
%H=0.25*(diag(1./mux)*psq+psq*diag(1./mux));%Eq.(4)
%H=0.5*diag(1./mux)*psq;                    %Eq.(5)
%H=0.5*psq*diag(1./mux);                    %Eq.(6)

% add potential part
Vpot=polyval(fliplr(K),(x*aBohr).^2); % evaluate
                                      % polynomial
H=H+diag(Vpot);
\end{verbatim}
\end{small}
Eigenenergies and eigenfunctions can now be found using:
\begin{small}
\begin{verbatim}
[Wavefuncs,En]=eig(H);
\end{verbatim}
\end{small}
The following MATLAB \cite{gen:matl} program generates
the matrix representation for the two-dimensional 
Henon-Heiles system (\ref{eq:2DHHS}) using $N_x=N_y=N=2M+1$ points
with $L_x=L_y=L$:
\begin{small}
\begin{verbatim}
[x,ip_op] = gen_mommatrix(L,M);
psq=-ip_op^2;
del=eye(2*M+1); % unit matrix
H=0.5*(kron(psq,del)+kron(del,psq)+...
   kron(diag(x.^2),del)+kron(del,diag(x.^2)))+...
   (1/sqrt(80))*(kron(diag(x.^2),diag(x))-...
   kron(del,diag(x.^3))/3);
\end{verbatim}
\end{small}
We have used the MATLAB function {\tt kron} that builds the tensor (Kronecker)
product of two matrices.
The generalization to discretizations that are
anisotropic in $x$ and $y$ and also to more than two dimensions
is obvious, and eigenvalues of $\op{H}$ are found as before.
\end{appendix}

%\section*{References}
\bibliographystyle{apsrev}
%\bibliographystyle{unsrt}

%\bibliography{journals,gen,sfm,uw}

%\bibliography{bib/aies,bib/anti,bib/bibHelp,bib/bsp,bib/cntrl,bib/cold,bib/csm,bib/dia,bib/dr,bib/dtmu,bib/gen,bib/journals,bib/math,bib/muster,bib/nu,bib/o_flair,bib/priv,bib/qd,bib/qdm,bib/sct,bib/sfa,bib/sfm,bib/vdw,bib/zpc}
% \bibliography{/data/amo/group/tex/bib/sfm,
%              /data/amo/group/tex/bib/sfa
%              /data/amo/group/tex/bib/dia,
%              /data/amo/group/tex/bib/gen}

%\bibliography{sfm,sfa,dia,gen}
% \bibliography{../bibliography/bib/sfa,%
%               ../bibliography/bib/dia,%
%               ../bibliography/bib/sfm,%
%               ../bibliography/bib/gen,%
%               ../bibliography/work_02}

\end{document}